\documentclass[prl,twocolumn,tightenlines,showpacs,nofootinbib]{revtex4}
\usepackage{bm,dcolumn,amsmath,graphicx,amsfonts,amssymb}

\begin{document}
\title{Calculation of parity non-conserving optical rotation in iodine at 1315 nm}

\author{G. E. Katsoprinakis, L. Bougas}
\author{T. P. Rakitzis}
\email{ptr@iesl.forth.gr}
\affiliation{Institute of Electronic Structure and Lasers, Foundation for Research and Technology-Hellas, 71110 Heraklion-Crete, Greece,}
\affiliation{Department of Physics, University of Crete —71003 Heraklion-Crete, Greece}
\author{V. A. Dzuba}
\email{dzuba@phys.unsw.edu.au}
\author{V. V. Flambaum}
\affiliation{School of Physics, University of New South Wales, Sydney 2052,
Australia}

\date{ \today }

\begin{abstract}
We examine the feasibility of a parity non-conserving (PNC) optical rotation experiment for the $^2$P$_{3/2}\rightarrow ^2$P$_{1/2}$ transition of atomic iodine at 1315 nm. The calculated $E1_{\rm PNC}$ to $M1$ amplitude ratio is $R=0.80(16)\times 10^{-8}$. We show that very large PNC rotations (greater than 10 $\mu$rad) are obtained for iodine-atom column densities of $\sim 10^{22}$ cm$^{-2}$, which can be produced by increasing the effective interaction pathlength by a factor of $\sim 10^4$ with a high-finesse optical cavity. The simulated signals indicate that measurement of the nuclear anapole moment is feasible, and that a 1\% PNC precision measurement should resolve the inconsistency between previous measurements in Cs and Tl.
\end{abstract}
\pacs{11.30.Er; 12.15.Ji; 31.15.A-}
\maketitle

%%% remove comment delimiter ('%') and select language if required
%\selectlanguage{spanish} 

%\section{INTRODUCTION}
%
\indent\emph{Introduction - } The precise measurement of atomic parity nonconservation (PNC) provides a low-energy test of the standard model and of internucleon weak interactions \cite{PNCrev0,PNCrev1,PNCrev2}. In recent decades, successful atomic PNC measurements have been performed using two techniques: (a) the Stark interference technique, on Cs \cite{CsScience}, and (b) the optical rotation technique, on Tl, Bi, and Pb \cite{TlFortson,TlEdwards,Bi,Pb}. The highlight of these efforts was the 0.35\% precision measurement of nuclear-spin-independent PNC in Cs, and the 14\% precision measurement of the nuclear spin-dependent PNC for the odd-proton nucleus of ${}^{133}$Cs \cite{CsScience}. However, measurement of the anapole moment in Cs disagrees with Tl \cite{TlFortson,CsScience}, and also with some theoretical nuclear calculations (\cite{GingesFlambaum2004,BouchiatPiketty,HaxtonWieman2001} and references therein). To help resolve these inconsistencies, and to improve the atomic PNC tests of the standard model, further experiments are needed. For
example, there is a proposal to measure the nuclear anapole moment using the PNC-generated hyperfine frequency shift of dressed states in atoms \cite{FS1993}, with particular proposals for Cs \cite{BouchiatCs} and Fr \cite{BouchiatFr}. However, for other PNC candidates, as the precision in the atomic theory is not expected to significantly surpass the current experimental or theoretical PNC precision of Cs, future PNC experiments have focused on other fruitful directions, such as the measurement of atomic PNC on a chain of isotopes \cite{isotopes,GingesFlambaum2004}, or the measurement of nuclear anapole moments.  Therefore, along these lines, PNC experiments are in progress on Yb and Dy at Berkeley \cite{YbBerkeley,Dy}, on Rb and Fr at the Tri-University Meson Facility (TRIUMF, University of British Columbia) \cite{RbFr1,RbFr2}, on Ra$^{+}$ at KVI Groningen \cite{Ra}, and on Ba${}^{+}$ at the University of Washington, Seattle \cite{Ba}.
Recently, Bougas \emph{et al.} proposed the measurement of optical rotation in excited (metastable) states of Hg and Xe \cite{prl}, using a novel optical bow-tie cavity to enhance available single-pass metastable column densities from $10^{14}$ to $10^{18}$ cm$^{-3}$ (by achieving $\sim$10$^{4}$ cavity passes) and to effect two new signal reversals. Hg and Xe both have several stable isotopes, and each have two isotopes with odd-neutron nuclei (for which anapole moments have not yet been measured). It will be useful to find compatible atomic PNC candidates that allow measurement of the anapole moment of odd-proton nuclei as well, needed to resolve the inconsistencies in the Cs and Tl measurements. Here we investigate the suitability of ground-state $^{127}$I (which has an odd-proton nucleus) as a candidate for measuring nuclear spin-dependent PNC. Note that although $^{127}$I is the only stable isotope of iodine, several radioactive isotopes are also commercially available. Therefore, the aim of this paper is to present spin-independent and spin-dependent PNC calculations for the $^{2}$P$_{3/2}\rightarrow ^{2}$P$_{1/2}$ spin-orbit transition of $^{127}$I and to examine the experimental feasibility of PNC measurements for this transition using the cavity-enhanced optical rotation technique.
%
%\section{Calculations}
%
\newline
\newline\indent\emph{Calculations - }
The Hamiltonian describing parity-nonconserving electron-nuclear
interaction can be written as a sum of the nuclear-spin-independent (SI) and
the nuclear-spin-dependent (SD) parts (we use atomic units: $\hbar = |e| = m_e = 1$):
%------------------------------------------------------------------
\begin{eqnarray}
     H_{\rm PNC} &=& H_{\rm SI} + H_{\rm SD} \nonumber \\
      &=& \frac{G_F}{\sqrt{2}}                             %  (1)
     \Bigl(-\frac{Q_W}{2} \gamma_5 + \frac{\varkappa}{I}
     {\bm \alpha} {\bm I} \Bigr) \rho({\bm r}),
\label{e1}
\end{eqnarray}
%------------------------------------------------------------------
where  $G_F \approx 2.2225 \times 10^{-14}$ a.u. is the Fermi constant of
the weak interaction, $Q_W$ is the nuclear weak charge,
$\bm\alpha=\left(
\begin{array}
[c]{cc}%
0 & \bm\sigma\\
\bm\sigma & 0
\end{array}
\right)$ and $\gamma_5$ are the Dirac matrices, $\bm I$ is the
nuclear spin, and $\rho({\bf r})$ is the nuclear density normalized to 1.
Within the standard model
the weak nuclear charge $Q_W$ is given by~\cite{SM}
\begin{equation}
Q_W \approx -0.9877N + 0.0716Z.
\label{eq:qw}
\end{equation}
Here $N$ is the number of neutrons, and $Z$ is the number of protons.
The strength of the spin-dependent PNC interaction is proportional to
the dimensionless constant $\varkappa$, which is to be found from the
measurements. It is believed that $\varkappa$ is dominated by the
nuclear anapole moment~\cite{anapole,anapole2}.
The PNC amplitude of an electric dipole transition between states of
the same parity $|i\rangle$ and $|f \rangle$ is equal to
\begin{eqnarray}
   E1^{\rm PNC}_{fi}  &=&  \sum_{n} \left[
\frac{\langle f | {\bm d} | n  \rangle
      \langle n | H_{\rm PNC} | i \rangle}{E_i - E_n}\right.
\nonumber \\
      &+&
\left.\frac{\langle f | H_{\rm PNC} | n  \rangle
      \langle n | {\bm d} | i \rangle}{E_f - E_n} \right],
\label{eq:e2}
\end{eqnarray}
where ${\bm d} = -e\sum_i {\bm r_i}$ is the electric dipole operator.
To extract from the measurements the parameter of the nuclear
spin-dependent weak interaction, $\varkappa$, one needs to consider PNC
amplitudes between specific hyperfine structure components of the
initial and final states. The detailed expressions for these
amplitudes can be found in earlier works~\cite{Porsev01,JSS03,Xe}.
We perform the calculations of the PNC amplitude using the
configuration interaction (CI) method.  The PNC amplitude
(\ref{eq:e2}) in iodine is dominated by the $5s$ -- $5p$ transition.
Therefore we treat $5s$,
$5p_{1/2}$, and $5p_{3/2}$ states as valence states, performing CI
calculations for seven valence electrons. Calculations are done in the
$V^{N}$ approximation, which means that the initial Hartree-Fock
procedure is done for neutral iodine. The hole in the $5p$ subshell is
taken into account via the fractional occupation number
techniques~\cite{DFM03,DF08a,DF08b}. 
The complete set of single-electron orbitals is
constructed using the B-spline technique~\cite{JohSap86}. We calculate
40 B splines in the cavity of $R=40a_{\rm B}$ in each partial wave up to
$l_{\rm max}=2$. We use 14 lowest states above the core in each
of the $s$, $p_{1/2}$, $p_{3/2}$, $d_{3/2}$, and $d_{5/2}$ in the CI
calculations. Core-valence correlations are not included because in a
many-valence-electron system such as iodine, intervalence correlations
strongly dominate over the core-valence correlations. 
The effects of external fields, such as the electric field of the laser light
and electron-nucleus parity-nonconserving weak interactions, are taken
into account within the framework of the random-phase approximation (RPA) 
(see, e.g., \cite{CPM}). The summation over intermediate many-electron
states in (\ref{eq:e2}) is done with the use of the Dalgarno-Lewis
technique~\cite{DalLew55} (see also \cite{Xe,Yb}).
The main factor contributing to the uncertainty of present calculations is
the incompleteness of the many-electron basis. This is very typical for
the CI calculations with a large number of valence electrons. Full
saturation of the basis would lead to a matrix of unmanageable
size. To reduce the size of the matrix, we cut our basis in such a way
that all important configurations are included. First, we limit our
single-electron basis to $l_{\rm max}=2$. Second, we consider only single
and double excitations when we construct basis configurations from the
ground-state configuration. Then, we limit single excitations to 14
lowest states above the core. Usually 14 out of 40 B splines is
enough to saturate single-electron excitations. Then we limit
double excitations to the eight lowest states above the core. 
Comparing different runs of the code with different cuts to the basis
we estimate that the uncertainty of the present calculations of the PNC
amplitude is about 20\%. This uncertainty can be further reduced in
case of progress with the measurements by using more computer power.
The resulting PNC amplitude ($z$ component) is 
\begin{equation}
 E1_{\rm PNC} = 1.46(29) \times 10^{-11} (-Q_W/N) iea_{\rm B}.
\label{eq:pnc}
\end{equation}
Using \eqref{eq:qw} and \eqref{eq:pnc} we obtain the value of the reduced matrix
element for the $E1_{\rm PNC}$ transition amplitude between states
$^2$P$^{\rm o}_{3/2}$ and $^2$P$^{\rm o}_{1/2}$ of $^{127}$I:
\begin{equation}
 E1_{\rm PNC} = 3.35(67) \times 10^{-11} iea_{\rm B}.
\label{eq:pncrme}
\end{equation}
The spin-dependent PNC amplitudes between different
hyperfine-structure (hfs) components of the ground and excited states
are presented in Table~\ref{t:pnc}.
The $M1$ amplitude between the $^2$P$^{\rm o}_{3/2}$ and $^2$P$^{\rm o}_{1/2}$
states is $0.00420 \ ea_{\rm B}$. Here we took into account 
that the overlap of the $5p_{1/2}$ and $5p_{3/2}$ functions is found from the Hartree-Fock calculations to be 
$\langle 5p_{1/2}\mid 5p_{3/2}\rangle =$0.997. The uncertainty for the M1 amplitude is order 0.1\%. The angle of the optical rotation is proportional to the ratio $R={\rm Im}(E_{\rm PNC}/M1)$. Using (\ref{eq:pncrme}) and $M1=0.0042 ea_{\rm B}$
we get $R=0.80(16) \times 10^{-8}$. Note that the dependence of the $E_{\rm PNC}$  
and $M1$ amplitudes on the values of the total angular momentum $F$ (including nuclear spin) and
its projection $M$ is the same if we neglect the $\varkappa$ contribution to
$E_{\rm PNC}$ and the nuclear magnetic moment contribution to $M1$. Therefore,
for the nuclear-spin-independent contributions the values of $R$ are the
same for all transitions in Table \ref{t:pnc}.
%
%Note that the dependence of the
%$E1_{\rm PNC}$ and $M1$ amplitudes on the values of the total angular
%momentum $F$ (including nuclear spin) and its projection $M$ is the
%same. Therefore, the values of $R$ are the same for all transitions in
%Table~\ref{t:pnc}. 
%
\begin{table}
\caption{PNC amplitudes ($z$ components) for the $|5p^5 \ ^2$P$^{\rm o}_{3/2},F_1,M_{\rm max} \rangle \rightarrow  |5p^5 \ ^2$P$^{\rm o}_{1/2},F_2,M_{\rm max}\rangle$ ($M_{\rm max} = \min\{F_1,F_2\}$) transitions in $^{127}$I. ${\bm F} = {\bm J}+{\bm I}$, where ${\bm I}$ is nuclear spin, $I=5/2$.}
\label{t:pnc}
\begin{ruledtabular}
\begin{tabular}{ccc}
$F_1$ & $F_2$ & PNC amplitude $[10^{-11}iea_{\rm B}]$ \\
\hline
 1 & 2 & $0.92(18)[1 - 0.043(9) \varkappa]$ \\
 2 & 2 & $1.21(24)[1 - 0.045(9) \varkappa]$ \\
 2 & 3 & $0.39(8) [1 + 0.037(7) \varkappa]$ \\
 3 & 2 &$-0.64(13)[1 - 0.048(10)\varkappa]$ \\
 3 & 3 & $1.08(22)[1 + 0.035(7) \varkappa]$ \\
 4 & 3 &$-0.84(17)[1 + 0.031(6) \varkappa]$ \\
\end{tabular}
\end{ruledtabular}
\end{table}
%
%\section{EXPERIMENTAL FEASIBILITY}
%
\newline
\newline\indent\emph{Experimental feasibility - }
PNC optical rotation is given by 
\begin{equation}
\varphi _{\rm PNC} =-\frac{4\pi \; {l}}{\lambda } \left[n(\omega )-1\right]R 
\end{equation} 
where $l$  is the length of vapor, $\lambda$ is the optical wavelength, $\omega$ is the optical frequency, and $n(\omega)$ is the refractive index due to the absorption line. For optical depths $L\ll$ 1, the optical rotation angle ${\varphi}_{\rm PNC} \propto \rho l$ (where $\rho$ is the density and $\rho l$ is column density of the vapor). For optical depths $L\gg$ 1, the vapor is optically thick near the line center where ${\varphi}_{\rm PNC}$ is largest and can no longer be observed. The effective maximal rotation angle is shifted further off resonance as $\sqrt{\rho l}$, and $\varphi _{\rm PNC}^{\max } \propto \sqrt{\rho l}$, as can be shown by maximizing the product of dispersion and transmission. Therefore, the rotation angle can still be increased with increasing column density for L $\gg$ 1. For this reason, in the PNC optical rotation experiments using Tl, Bi, and Pb vapors, optical depths of about 50 were used (corresponding to column densities of $\sim 10^{19}$ cm$^{-2}$) to obtain optical rotations of $\sim$1 $\mu$rad.
The value of $R = 0.8 \times 10^{-8}$ for $^{127}$I is about an order of magnitude smaller than that for Tl, Bi, and Pb (whereas the $M1$ amplitude and transition wavelength $\lambda$ are about the same in all four cases, and the line-shape parameters are approximately similar). Therefore to compensate for the smaller value of $R$ and to produce PNC optical rotations of $\sim$1 $\mu$rad for $^{127}$I, larger optical depths of $\sim 10^3$ are necessary (corresponding to column densities of order $\sim 3\times {10}^{20}$ cm$^{-2}$). Such a large column density can realistically be achieved only with the use a high-finesse cavity. We first discuss the feasibility of producing column densities up to $\sim$10$^{22}$ cm$^{-2}$, and then use such column densities in simulations of optical rotation profiles for $^{127}$I.
%
%\subsection{Production of high density I($^2$P$_{3/2}$) atoms}
%
\newline\indent

High iodine atom densities of $\sim 10^{16}$ cm$^{-3}$ have been achieved in dc glow discharges \cite{Idisch1,Idisch2,Idisch3}. However, in achieving these densities through discharge, relatively high pressures ($\sim$few tens of Torr) of both the precursor (I$_2$, CH$_3$I) and carrier gases (He, Ar) are required. Such pressures can significantly suppress the measured PNC signal, through the induced collisional broadening of the $^2$P$_{3/2}\rightarrow^2$P$_{1/2}$ transition. Therefore, we consider an alternative method for the creation of ground-state $^2$P$_{3/2}$ iodine atoms at lower pressures through photodissociation of I$_2$ molecules:
\begin{equation}\label{eq:I2pd}
{\rm I}_{2}\stackrel{h\nu }{\longrightarrow} {\rm I}(^{2}{\rm P}_{3/2}) + {\rm I}(^{2}{\rm P}_{3/2})
\end{equation}
%
%\noindent The absorption cross section of I$_{2}$ at 308 K is shown in Fig. \ref{fig:I2abs} \cite{I2abs}. We see that the peak value is at $\sim$530 nm with a value of $\sigma(530 {\rm nm}) = 2.8 \times 10^{-18}$ cm$^{2}$. The production of spin-orbit-excited I($^{2}$P$_{1/2}$) atoms is energetically allowed only for photodissociation wavelengths less than $\sim$500 nm:
%
\begin{figure}
	\includegraphics[width=\linewidth]{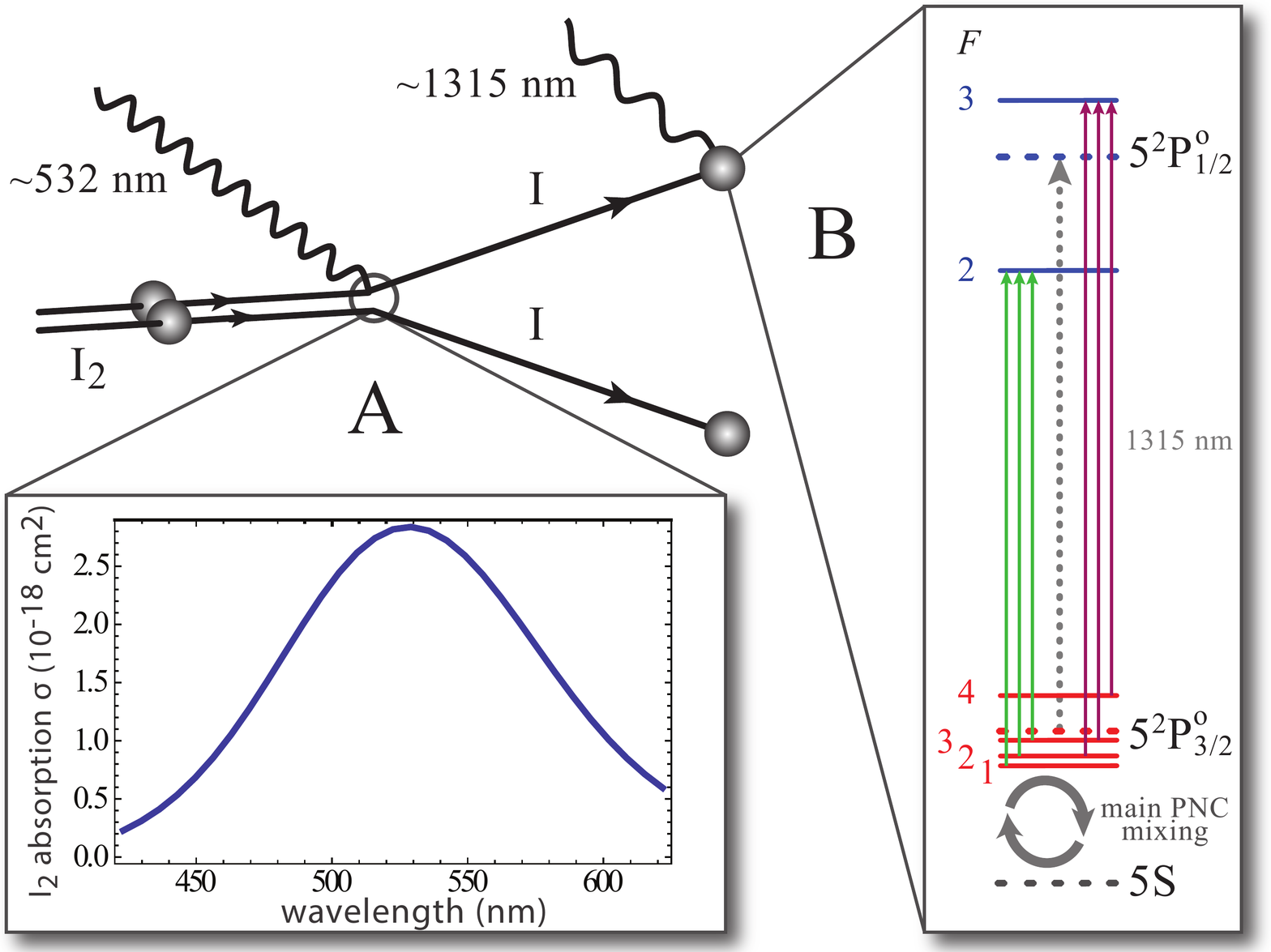}
	\caption{\label{fig:I2abs} (A) Photodissociation of I$_2$ (ideally with light at $\sim$532 nm) and the corresponding absorption cross section (taken from \cite{I2abs}). (B) Level scheme for the $^2$P$_{3/2}^{\rm o}\rightarrow^2$P$_{1/2}^{\rm o}$ $M$1 transition of atomic iodine at 1315 nm. Indicated in gray color are the individual $F\rightarrow F'$ transitions constituting the two separated hyperfine groups of Fig. \ref{fig:I2PNCVarLorDen}. The main contribution to PNC mixing comes from the 5$S$ core states.}
\end{figure}
\noindent The absorption cross section of I$_{2}$ at 308 K peaks at $\sigma = 2.8 \times 10^{-18}$ cm$^2$ for $\lambda\sim$530 nm (see Fig. \ref{fig:I2abs} and \cite{I2abs}). The production of spin-orbit-excited I($^{2}$P$_{1/2}$) atoms:
\begin{equation}\label{eq:I2pdforb}
{\rm I}_{2}\stackrel{h\nu }{\longrightarrow} {\rm I}(^{2}{\rm P}_{3/2}) + {\rm I}(^{2}{\rm P}_{1/2}),
\end{equation}
\noindent is energetically allowed only for photodissociation wavelengths smaller than $\sim$500 nm, so this undesirable channel is avoided by photodissociating the I$_2$ molecules at higher wavelengths, e.g., at 532 nm, with the second harmonic of a Nd:YAG laser or with a green light-emitting diode array. At room temperature, the vapor pressure of solid I$_{2}$ yields a density of about 5 $\times$ 10$^{15}$ cm$^{-3}$ I$_{2}$ vapor and a column density (over a 100-cm path length) of 5 $\times$ 10$^{17}$ cm$^{-2}$. This corresponds to an optical depth of $\sim$1.
The I($^{2}$P$_{3/2}$) atoms recombine with the three-body reaction:
\begin{equation}\label{eq:Irec}
{\rm I}(^{2}{\rm P}_{3/2}) + {\rm I}(^{2}{\rm P}_{3/2}) +{\rm I}_{2}\stackrel{k_{r} }{\longrightarrow} 2 {\rm I}_{2},
\end{equation}
\noindent where the recombination rate is given by $k_{r}= 3.1(3) \times 10^{-30}$ cm$^{2}$ molecule$^{-2}$ s$^{-1}$ at 308 K \cite{I2rec}. Setting the photodissociation rate and recombination rate of I$_{2}$ to be equal, we obtain: 
\begin{equation} \label{eq:I2rateeq} 
\frac{d[{\rm I}_{2} ]}{dt} =-J[{\rm I}_{2} ]+k_{r} [{\rm I}_{2} ][{\rm I}]^{2} =0,
\end{equation} 
\noindent where $J$ is the I$_2$ photodissociation rate constant, given by $J = \Phi\sigma$, and $\Phi$ is the photon flux. Solving for the atomic iodine concentration yields [I] = $\sqrt{\Phi\sigma/k_r}$. Choosing $\Phi\sim$ 10$^{20}$ photons cm$^{-2}$ s$^{-1}$ (corresponding to 50 W cm$^{-2}$ at 532 nm, which is probably close to a reasonable upper bound) gives a steady-state iodine-atom concentration [I] $\sim$ 10$^{16}$ cm$^{-3}$. At I densities of 10$^{16}$ cm$^{-3}$, the mean time between wall collisions (for a cell-tube diameter of about 1 cm) is $\sim$1 ms. The sticking probability of I atoms with the cell walls is typically on the order of 5\% but can be reduced to 10$^{-3}$ for acid-coated cells \cite{Iwall,Iwall2}, so that loss of I atoms from sticking and recombination at the walls should occur on the $\sim$1-s time scale. The I atoms should collide with the cell walls hundreds of times on average, ensuring that they are thermalized. Notice that we calculate a steady-state concentration [I$_2$] = 0 (the initial [I$_{2}$]$_{\rm o}$ = 5 $\times$ 10$^{15}$ cm$^{-3}$ has been photodissociated to give [I] = 10$^{16}$ cm$^{-3}$), as we have ignored recombination processes with smaller rates in Eq. \eqref{eq:I2rateeq}, such as I-atom recombination at the walls and three-I-atom recombination. However, as these reactions are expected to be less significant than \eqref{eq:Irec} at these pressures, the final I$_{2}$ concentration can be kept at $\sim$10$^{15}$ cm$^{-3}$. Finally, using a cell with a length of 100 cm and a high-finesse optical cavity with $\sim$10$^{4}$ passes gives an upper bound for the effective I-atom vapor column density of 10$^{22}$ cm$^{-2}$.
%
%\subsection{Simulations of PNC optical rotation}\label{ssec:sim}
%
\newline\indent
\begin{figure}[h!]
	\includegraphics[width=\linewidth]{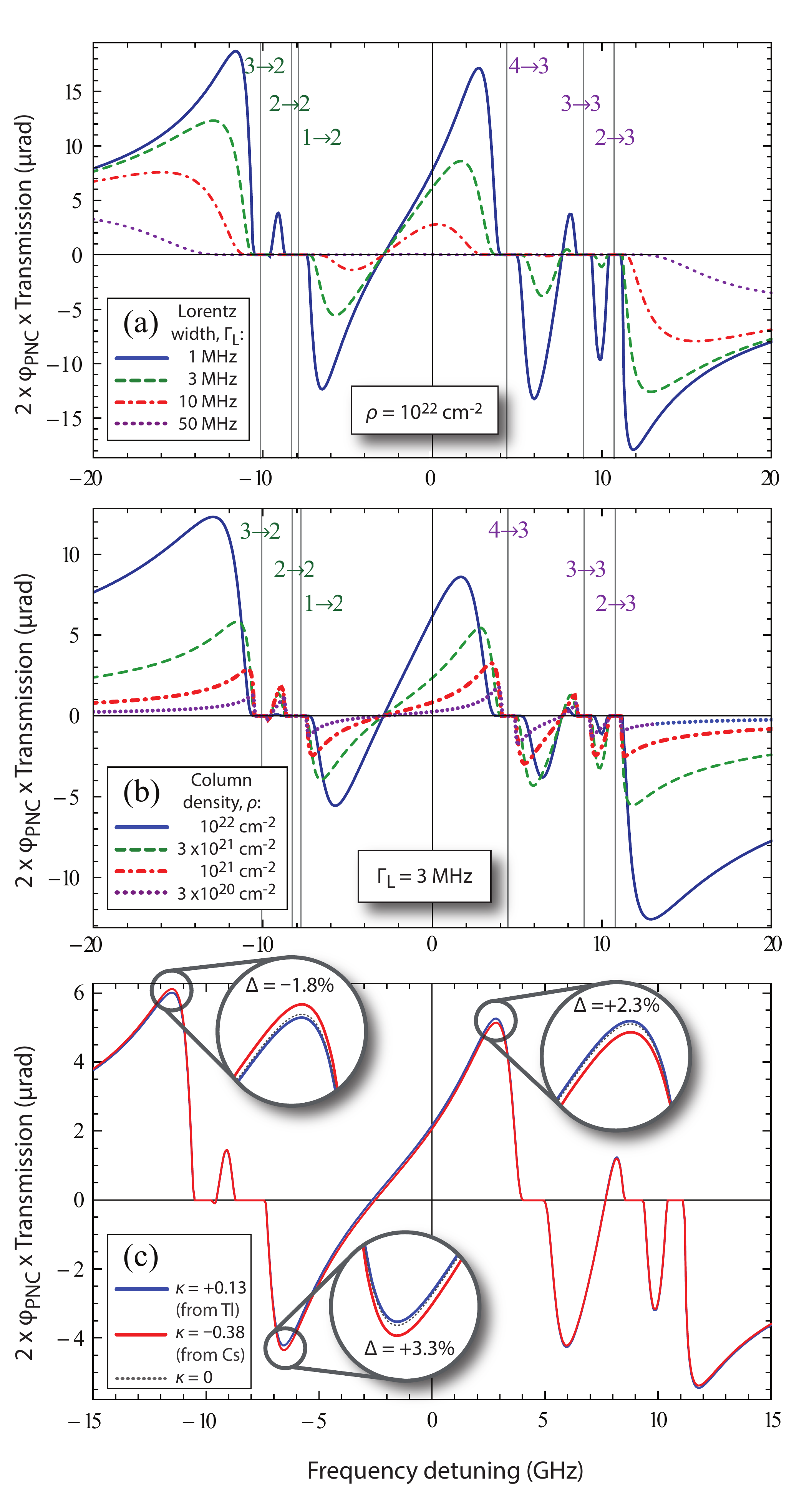}
	\caption{\label{fig:I2PNCVarLorDen} Simulated signal vs frequency. (a) For the same column density ($10^{22}$ cm$^{-2}$) and various Lorentz contributions (from 1 to 50 MHz). (b) For the same Lorentzian width (3 MHz) and various column densities (from 10$^{22}$ down to 3$\times$10$^{20}$ cm$^{-2}$). (c) Effect of the nuclear anapole moment. This is the simulated signal for $\rho=3\times 10^{21}$ cm$^{-2}$ and $\Gamma_{\rm L}=$3 MHz, for $\varkappa=0.13$ and $\varkappa=-0.38$. For a discussion, see text.}
\end{figure}
In a typical PNC optical rotation experiment, the observable is the product of $\varphi_{\rm PNC}$ times the transmitted optical power and this is doubled by the signal reversals proposed in \cite{prl}.
%the potential for rapid alternation between the polarization mode pairs, resulting either from the swift reversal of the magnetic field or from switching the laser frequency between the two mode pair resonances, leads to the doubling of this typical rotation signal.
Simulations of PNC optical rotation in the $^{2}$P$_{3/2}\rightarrow ^{2}$P$_{1/2}$ $M$1 transition in iodine are shown in Fig. \ref{fig:I2PNCVarLorDen}. In Fig. \ref{fig:I2PNCVarLorDen}(a) we assume a column density of 10$^{22}$ cm$^{-2}$ and the dependence on the Lorentzian broadening of the line is investigated. We see that an increasing Lorentz width reduces the measured PNC signal. As the Lorentz width is mainly determined by the pressure inside the cell, maximizing the ratio of iodine-atom pressure to total pressure is desirable. The collision-broadened Lorentz widths for the $M$1 transition for I and I$_2$ collision partners were estimated in \cite{ILor} to be 4.3 and 10.3 MHz/Torr at room temperature, respectively. We see that, for realistic Lorentz widths, signals in excess of 10 $\mu$rad can be achieved, along with sufficient hyperfine resolution, which is necessary in determining the $\varkappa$-value of Table \ref{t:pnc}.
\newline\indent In Fig. \ref{fig:I2PNCVarLorDen}(b) the reverse situation is examined: a constant Lorentz width is maintained, determined by fixed densities of I and I$_2$, and a varying number of passes is assumed, ranging from 10$^4$ down to 300. As we can see, even in the case of 300 passes, a signal on the order of $\sim$1 $\mu$rad at the peak is predicted (similar to the signal levels measured for Tl \cite{TlFortson}), allowing for a significant margin of error in the experimental constraints. 
\newline\indent Finally, in Fig. \ref{fig:I2PNCVarLorDen}(c), the nuclear spin-dependent PNC effects are examined. We use the previously measured values of $\varkappa$ from the thallium \cite{TlFortson} and cesium \cite{CsScience} experiments to predict values for $^{127}{\rm I}$. Using the simple single-valence-proton model \cite{anapole2}, we obtain:

%\begin{alignat}
%\approxeq(^{127}{\rm I})~ &\simeq  &\varkappa(^{205}{\rm Tl}) &\simeq & 0.13(20)  & {\rm [from~Tl]} \\
%\varkappa(^{127}{\rm I})~ &\simeq  -&\varkappa(^{133}{\rm Cs}) &\simeq -& 0.38(6) & {\rm [from~Cs]}
%\end{alignat}

\begin{alignat}{3}
\varkappa(^{127}{\rm I})~ & \simeq & \varkappa(^{205}{\rm Tl}) \;\;& \simeq & ~0.13(20) & \quad\text{[from Tl]}\\
\varkappa(^{127}{\rm I})~ & \simeq &~ -\varkappa(^{133}{\rm Cs}) & \simeq & ~-0.38(6) & \quad\text{[from Cs]}
\end{alignat}

\noindent These two values define the anticipated physical range of the spin-dependent part of the measurement and are the ones used for the simulations. We see that peak signal values differ by about $\pm$2-3\% for the two cases, resulting in total signal differences of up to  $\sim$5\%, as expected from the $\varkappa$ coefficients in Table \ref{t:pnc}. Therefore an experimental precision of 1\% or better is needed to clearly distinguish between these two extremes. Also evident from Table \ref{t:pnc} and reflected in Fig. \ref{fig:I2PNCVarLorDen}(c) is the fact that transitions belonging to the two separated hyperfine groups ($F\rightarrow F^\prime =2$ and $F\rightarrow F^\prime =3$) deviate in opposite directions in the two cases, meaning that a nuclear spin-dependent signal could be acquired even in the case where broadening mechanisms would blur the ground-state hyperfine structure. 

We note that within the same single-valence-proton model, $\varkappa(^{129}{\rm I}) \simeq \varkappa(^{131}{\rm I})  \simeq \varkappa(^{133}{\rm Cs})$, so that measurements on these iodine isotopes could provide additional experimental tests. The nuclear magnetic moments of these three nuclei (all with nuclear spin $I=\,^7/_2$) agree within 2\% \cite{nmmI129,nmmI131,nmmCs133}, which supports the validity of the single-valence-proton model. 
%
%\section{Conclusions}
%
\newline
\newline\indent\emph{Conclusions - }
Calculations for atomic iodine presented here along with signal simulations, indicate that a PNC optical rotation experiment is feasible, utilizing the cavity-enhanced, optical-rotation scheme of \cite{prl}. A nuclear anapole moment measurement seems possible, as the simulated signal levels and hyperfine resolution (shown in Fig. \ref{fig:I2PNCVarLorDen}) and the calculated amplitudes of nuclear spin-dependent PNC (shown in Table \ref{t:pnc}) compare favorably to those of the successful PNC optical-rotation experiment in thallium \cite{TlFortson}.
\acknowledgements
The work was supported by the European Research Council (ERC) grant
TRICEPS (GA No. 207542), by the National Strategic
Reference Framework (NSRF) grant Heracleitus II
(MIS 349309-PE1.30) co-financed by EU (European
Social Fund) and Greek national funds, and by the Australian Research Council.

\end{document}